# Random to chaotic temperature transition in low-field Fano-Feshbach resonances of cold thulium atoms

V.A. Khlebnikov[1], D.A. Pershin[1,2], V.V. Tsyganok[1,2], E.T. Davletov[1,2], I.S. Cojocaru[1,6], E.S. Fedorova[1,3], A.A. Buchachenko[4,5], A.V. Akimov[1,3,6]

[1]Russian Quantum Center, Business center "Ural", 100A Novaya str., Skolkovo, Moscow, 143025, Russia
[2]Moscow Institute of Physics and Technology, Institutskii per. 9, Dolgoprudny, Moscow region, 141700, Russia
[3]PN Lebedev Institute RAS, Leninsky prospekt 53, Moscow, 119991, Russia
[4]Skolkovo Institute of Science and Technology, Skolkovo Innovation Center, Moscow 143026, Russia
[5]Institute of Problems of Chemical Physics RAS, Chernogolovka, Moscow Region 142432, Russia
[6]Texas A&M University, 4242 TAMU, College Station, Texas, 77843, USA

e-mail: akimov@physics.tamu.edu

Here, we report on the observation of a random to chaotic temperature transition in the spacing of Fano-Feshbach resonances in the ultracold polarized gas of thulium atoms. This transition is due to the appearance of so-called $d$-resonances, which are not accessible at low temperatures, in the spectra at high temperatures, which drastically changes thulium's overall resonance statistic. In addition to this statistical change, it has been observed that $s$- and $d$-resonances experience quite different temperature shifts: $s$-resonances experience almost no shift with the temperature, while $d$-resonances experience an obvious positive shift. In addition, careful analysis of the broad Fano-Feshbach resonances enabled the determination of the sign of thulium's background scattering length. A rethermalization experiment made it possible to estimate a length value of $a_{bg} = +144 \pm 38$ a.u. . This proves that thulium atoms are suitable for achieving Bose-Einstein Condensation.

Ultracold dipolar atomic gases are becoming a powerful tool for quantum simulations. Rare-earth elements, having an incomplete $f$-shell, have a special place in this area because these atoms have a large orbital moment that comes from the orbital moment of electrons rather than their spin. This leads to a large number of low-field Fano-Feshbach resonances [1], enabling an extraordinary degree of control over interactions in these systems [2]. Recently chaotic spacing of Fano-Feshbach resonances was demonstrated in erbium and dysprosium [3] along with complicated temperature dependence of Fano-Feshbach resonances.

Ground-state thulium atoms have a total momentum of $F = 4$ with electron spin $S = 1/2$ and nuclear spin $I = 1/2$, thus having a 4 Bohr magneton magnetic moment in the ground state. Low spin quantum numbers make thulium level structure relatively simple compared to other lanthanides. This structure has convenient nearly-cycling cooling transitions [4,5]. Due to its high orbital moment thulium was expected to have many low-field Fano-Feshbach resonances similar to erbium and dysprosium [6–8]. However, thulium's smaller orbital momentum (having just one hole in its $f$-shell) should provide a reduction in the number of resonances compared to erbium and dysprosium and may lead to the absence of a chaotic statistic.

In this paper, we demonstrate that while a thulium atom has many $s$-resonances in a low magnetic field, it does not demonstrate chaotic behavior in resonance spacing at the low temperature limit. However, at higher temperatures (approximately 10 µK), at which $d$- and higher-order resonances appear in the spectrum, a chaotic statistic for the resonance spacing emerges. This transition occurs during a relatively small change in resonance density from $\rho \approx 3 G^{-1}$ to $\rho \approx 4.4 G^{-1}$. This transition behavior is quite different from erbium and dysprosium, in which chaotic statistics are temperature independent.

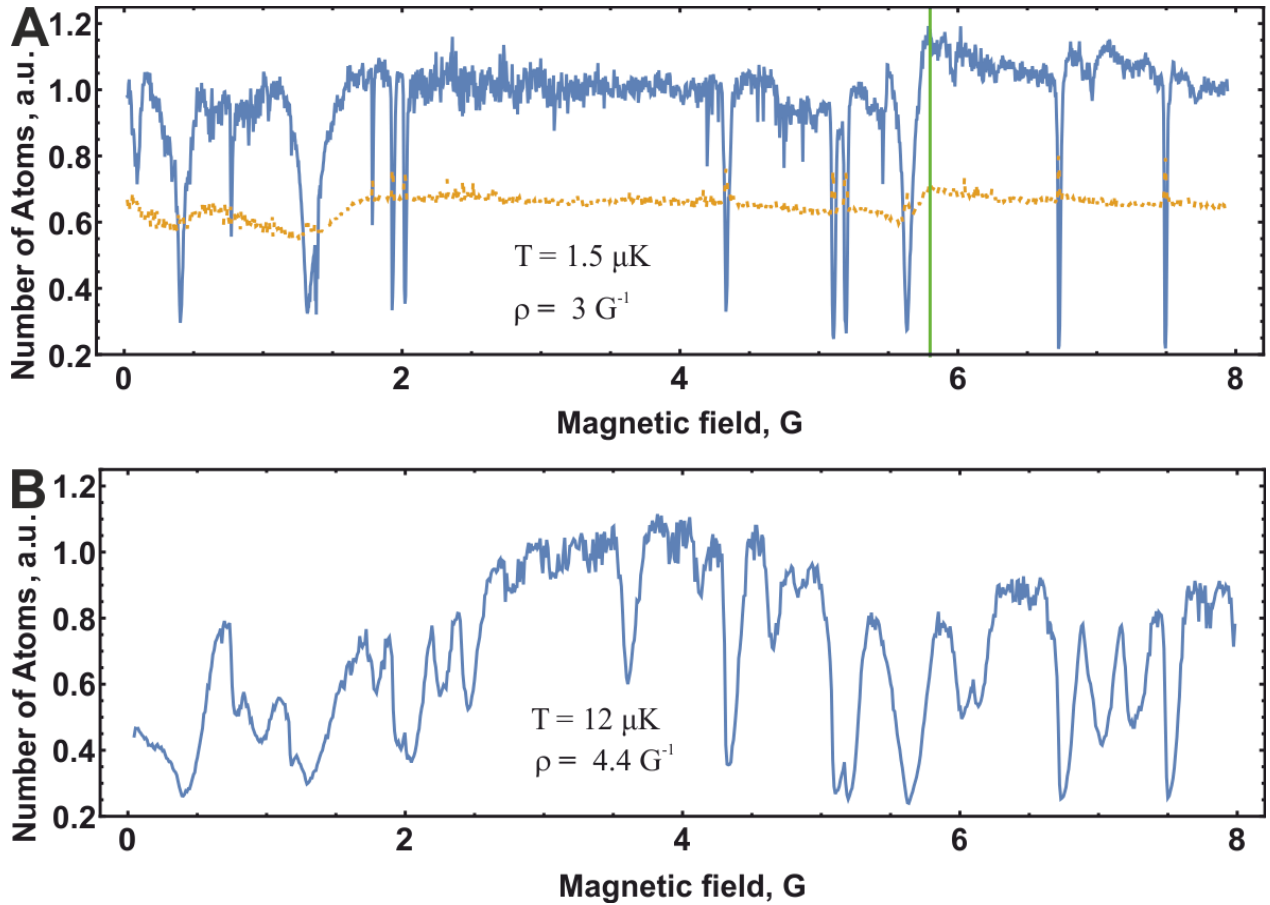

Figure 1 A – Detailed spectrum (5 mG step per point) of Fano-Feshbach resonances in the region 0-8G: the blue solid line represents the number of atoms versus magnetic field: the yellow dashed line marks the width of the atomic cloud after 0.5 ms of free expansion; the green vertical line at 5.8 G marks the magnetic field value at which scattering length vanishes. B – The same spectrum at 10 mG resolution and higher (12 $\mu$K) temperature.

The measured Fano-Fesbach resonance spectrum is presented at Figure 1A (see parallel PRA for experimental details). One can see reasonably dense spectra for resonances of various strengths covering the entire range of the magnetic field scan (8 G in this case). Sharp peaks in cloud size against the dips in atoms number in Figure 1A are most likely due to a low signal-to-noise ratio on the cloud's photographed at these fields. To understand the nature of the observed resonances we performed the same measurement but with less evaporative cooling, so that the temperature of the cloud was approximately 12 µK. One could see (Figure 1B) that all observed resonances split into two categories: either they broadened with the rising temperature, or they quickly grew from the background with the rising temperature.

The temperature dependence of the Fano-Feshbach resonance profiles of erbium and dysprosium at the temperatures below 2 µK were analyzed in [3] using the three-body recombination model for trap loss [9]. Following this model, we associate the quickly rising resonances with the $d$-type scattering Fano-Feshbach resonances, while those with modest intensity variation are associated with the $s$-type. Figure 2A and B shows the variations of the $s$-type resonance and $d$-type resonance at various magnetic fields

(see parallel PRA for more details). In agreement with the model [3], the intensity of the $s$-type decreases approximately as $(k_B T)^{-1}$, whereas that of the $d$-type increases as $k_B T$.

However, the observed resonance shifts strongly deviates from the linear model prediction [3]:

$$\Delta B = B - B_0 = (\lambda + 2)kT/\mu \tag{1}$$

where $\mu$ is the magnetic moment of the resonant trimer relative to that of the entrance channel, $B_0$ is the reference resonance shift defined here at the lowest attainable temperature, and $k_B$ is the Boltzmann constant, $\lambda = 0$ for the $s$-type resonances and $\lambda = 2$ for $d$-type resonances, respectively. As shown in the Figure 2A and B, the $s$-resonance centered at 4.4 G initially demonstrates a negligible shift, then a negative low field shift, accompanied by a sudden broadening. The shift of the $d$-resonance centered at 3.64 G continually increased, but with a nonlinear saturation trend. Similar behavior was observed for a few other resonances analyzed ($s$-type at 1.79, 1.95 and 2.04 G and $d$-type at 2.32, 2.51 Gauss).

For additional insight, we applied the same model in an attempt to simulate temperature dependence. In brief, the three-body loss rate coefficient is given by [3]:

$$L_3(T,B) \propto \frac{1}{2(k_B T)^3} \int_0^\infty d\varepsilon \varepsilon^2 \frac{\Gamma(\varepsilon)\Gamma_{br}}{[\varepsilon - \mu \Delta B]^2 + [\Gamma_{br} + \Gamma(\varepsilon)]^2/4} \exp(-\varepsilon/k_B T), \tag{2}$$

where $\varepsilon$ is the collision energy, $\Gamma_{br}$ and $\Gamma(\varepsilon) = A_\lambda \varepsilon^{\lambda+2}$ are the widths associated with the three-body resonance breakup and formation, respectively ($A_\lambda$ is $\lambda$-dependent coefficient of proportionality). Resonance shift (1) and $(k_B T)^{\lambda-1}$ intensity dependence comes from the formula in the narrow resonance limit [3]. We found that the model can qualitatively approximate the temperature dependence of the position and width (though not the intensity) of the $d$-type resonance simultaneously. It is known that the $L_3(T,B)$ rate saturates with the magnetic field detuning [10–12], and so does the whole resonance profile. This fact was confirmed by the simulations. Moreover, we even found the regime in which both shift and width start to decrease with $T$. Thus, the linear approximation (1) is valid at low temperatures when the overlap of the distorted Lorentzian and the Maxwell kinetic energy distribution is large enough. Still, it should be noted that we failed to attain complete and quantitative description of the measured profiles using $\Gamma_{br}$ and $A_\lambda$ as the only parameters of the model.

The observed behavior of the $s$-type resonance cannot be accounted for within the same model. Our assumption is that the three-body loss is already saturated, and other factors come into play on its background. The two-body loss mechanism is unlikely, which is determined by a similar model [13], except for a minor contribution from shallow-dimer – atom relaxation [14]. We speculate that the three-body shift toward a high magnetic field reduces the overlap with the magnetic profile of the true resonance given by the natural magnetic widths. This factor can be effectively accounted for by extra convolution and may also be responsible for the deviations seen in the $d$-resonance case. However, its elucidation would require the collection of data in a wider temperature range than is presently attainable.

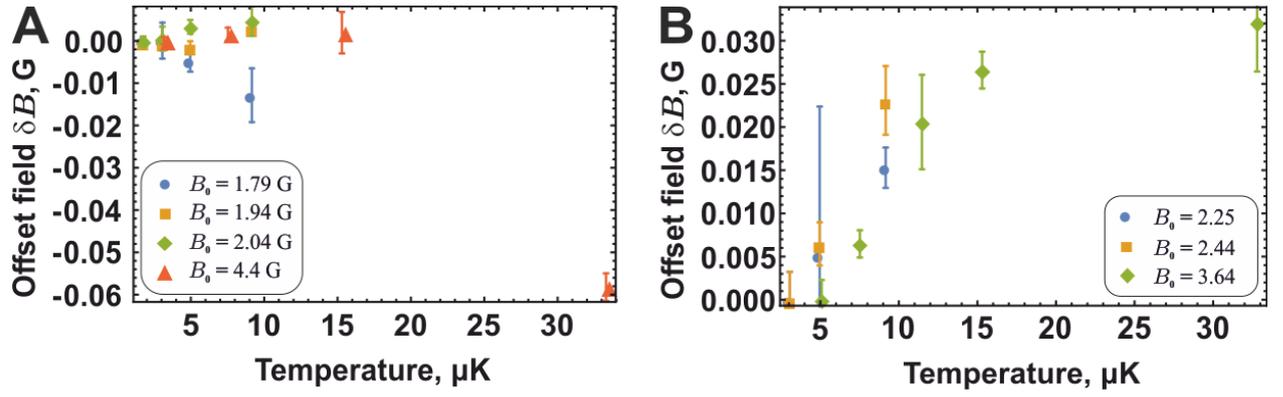

Figure 2 A – Shift of position of *s*-type Fano-Feshbach resonances versus temperature. Magnetic field, corresponding to the smallest temperature is indicated in the legend (see parallel PRA for details). B – Position of *d*-type Fano-Feshbach resonances versus temperature. Magnetic field, corresponding to the smallest temperature is indicated in the legend (see parallel PRA for details).

The statistic of measured $s$-resonances is presented in Figure 3A. Here, the resonance spacing $\delta B$ was normalized to average spacing between the resonances, which was found to be $\bar{\rho} = 2.97\, G^{-1}$, so normalized spacing is $s = \bar{\rho}\delta B$. We note that the average resonance spacing showed no significant deviation from linear behavior with the magnetic field (see inset to Figure 3B). It is observed that, probability distribution $P_P(s)$ and probability density $p_P(s)$ nicely follow an exponential distribution, which corresponds to independent resonance positions [3] and is typical for Poisson process.

Further insight into the statistic of the resonance spacing can be achieved via a graphical representation of the variance in the number of resonances versus the width of the magnetic field window $\Sigma^2(\Delta B)$ [3,13]. This variance is plotted in Figure 3B (see parallel PRA for details). Here, the expected statistic for Poissonian process is Poisson with average and variance equal to $\bar{\rho}\Delta B$. While $\Sigma^2$ shows correlations at rather large scales of the magnetic field (Figure 3B) the resonance spacing distribution allows the analysis of the behavior of resonances at short distances (Figure 3A). It is clear that both the resonance spacing and resonance number variance are close to the predictions of the random statistic.

At higher temperatures, as $d$-resonances become allowed, the statistic of Fano-Feshbach resonance spacing changes drastically, as is seen in Figure 3C. Here, spacing between the resonances begins to follow the Wigner-Dyson statistic [14]. This distribution appears when positions of resonances are not independent but are specified by some interaction potential of definite functional form. A comparison of experimental data and variances for Poisson and chaotic processes is presented in in Figure 3D. One can see that the high-temperature resonances tend to follow chaotic behavior. It is important to note that such resonance statistic behavior differs significantly from that for erbium, where the chaotic statistic is shown to be temperature independent [3].

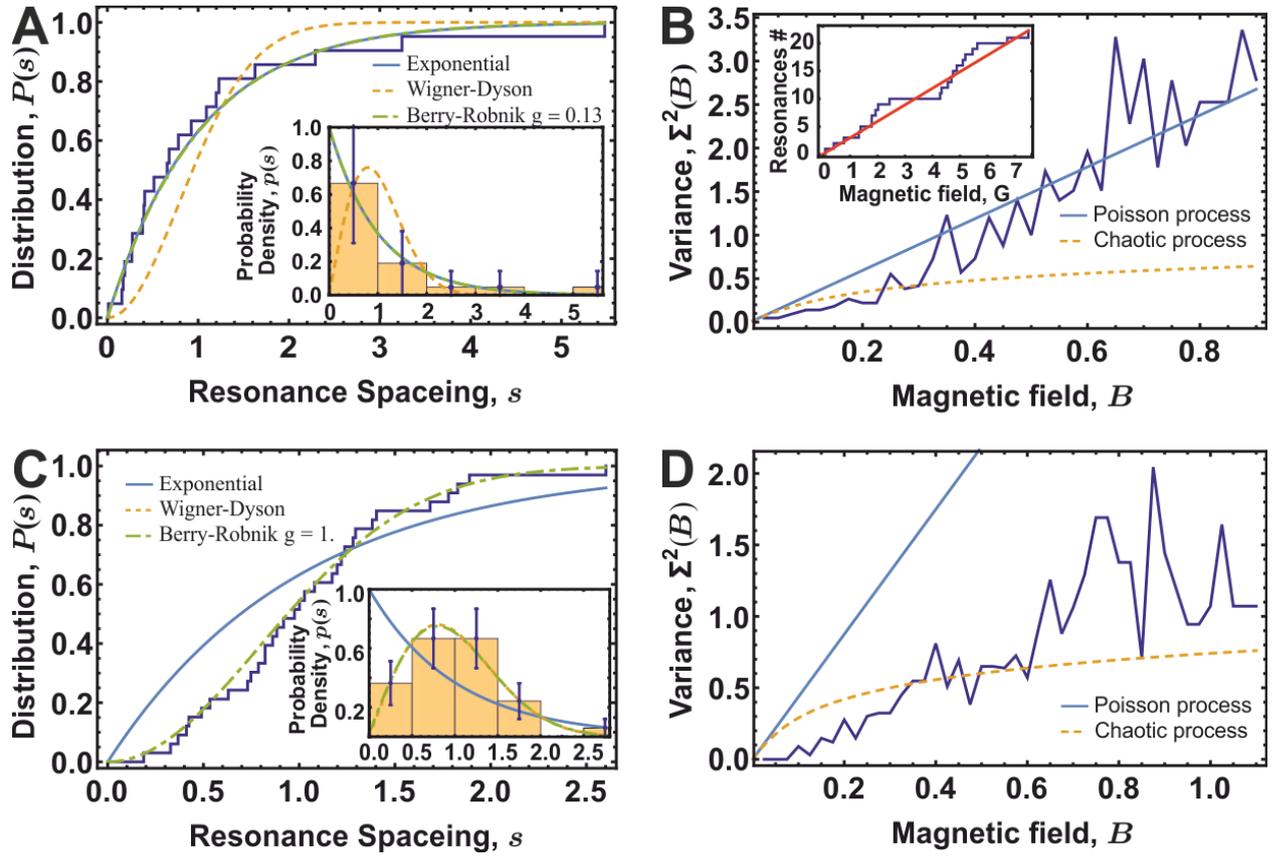

Figure 3 A – Probability distribution of normalized s-resonances spacing s for the first 8 Gauss of the magnetic field (resonances are presented in Figure 1A). Exponential and Berry-Robnik distributions overlap exactly. Inset shows the density distribution. The solid blue line represents Exponential distribution, the dashed orange line represents the Wigner-Dyson distribution; the dash-dot green line represents fit by Berry-Robnik distribution: Exponential and Berry-Robnik distributions overlap exactly. B – Variance of the number of s-resonances versus magnetic field window. C – Probability distribution of normalized s and d-resonances spacing s for the first 8 Gauss of the magnetic field (resonances are presented at Figure 1B). Inset shows the density distribution. Wigner-Dyson and Berry-Robnik distributions overlap exactly. D – Variance of the number of s- and d-resonances versus magnetic field window. Experimental points are calculated by binning resonances and calculating variance over the number of resonances in the bin. The minimum bin size was set to 0.025 G.

To further understand the statistic of Fano-Feshbach separations in more detail, we performed measurements of the resonances over a large range of magnetic fields at 1.5 $\mu K$. This way, only $s$-type resonances were observed. The average number of resonances per Gauss was found to be $\bar{\rho} = 2.31/\text{G}$ and is slightly lower than the range of 0-8 G, possibly due to a slightly lower resolution in the magnetic fields. While the statistic still tends to be random, both the probability distribution and variance tend to shift toward a chaotic statistic, much like what was predicted in [3] (Figure 4A). The degree of contribution of the chaotic and random statistic could be understood using a so-called Berry-Robnik distribution. The parameter $g$ gives a contribution of the chaotic statistic in the overall statistic, which is calculated as a sum of chaotic and random behaviors. One could see from Figure 4A, that while this statistic looks very close to a random one, the contribution of the chaotic statistic is quite significant. A comparison of the $s$-resonances in the 0-8 G range by the Berry-Robnik parameter was only 0.13, while for the sum of $s$- and $d$-resonances it was 1 (see Figure 3).

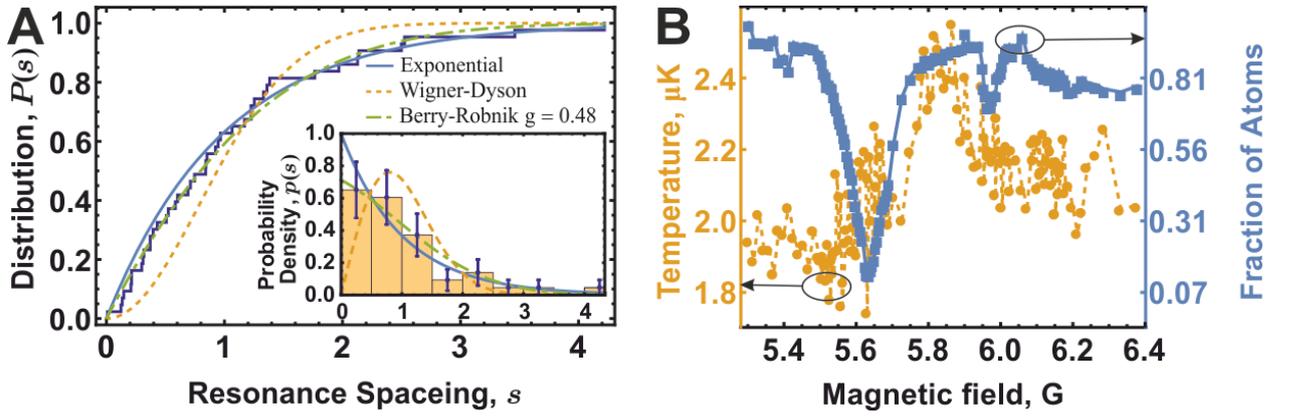

Figure 4 A – Statistic of the separation of Fano-Feshbach $s$ - resonances. Bin size corresponds to 300 mG. The blue solid curve represents exponential distribution, the orange dashed line represents the Wigner-Dyson distribution, and the green dot-dash line shows the Berry-Robnik distribution. The inset displays the density of the distribution. B – Variance of the number of $s$-resonances versus magnetic field window. Experimental points are calculated by binning resonances and calculating variance over the number of resonances in the bin. The minimum bin size was set to 0.05 G.

Wide Fano-Feshbach resonances also help in determining sign of background scattering length, which is of crucial importance for cooling down atomic ensemble to quantum degeneracy [15–17]. According to a description of resonance scattering with a quasi-bound intermediate state in s-wave, scattering length near resonance behaves as [1]:

$$a = a_{bg} - \frac{|a_{bg}\Gamma_0|}{\delta\mu(B-B_0)} \tag{3}$$

Here, $a_{bg}$ is the background scattering length in an open channel with no coupling to a closed channel, that is, the value of the scattering length far from resonance; $\Gamma_0$ is resonance strength and $\delta\mu$ is the difference between the magnetic moment of the separated atoms and the magnetic moment of the quasi-bound state. Some of the Fano-Feshbach resonances in Figure 1A have nonsymmetric behavior at the width of the atomic cloud, which is most clearly observed at the field values of 1.75 G and 5.83 G near the two broadest resonances. The change of the cloud's size on resonance could be due to different temperatures of the cloud around resonance or to fitting problems in the center of the cloud related to a greatly reduced number of atoms. To verify the source of this fluctuation, we performed a detailed temperature scan near the selected resonances, in which temperature was measured by a time-of-flight method. . While some resonances did not confirm a temperature change, the resonance peaked at 5.63 G (see Figure 4B) with a width of $\Delta = 0.2\,\text{G}$ and clearly has a temperature maximum on the right side of the resonance at the 5.83 G field. The closest $d$-type resonance lays well after the temperature resonance, and strong $s$-type resonance practically vanishes at the position of the temperature resonance. Given the proximity of Fano-Feshbach resonance, this maximum could be due to a vanishing scattering length and hence a minimizing elastic cross-section [8]. Indeed, atoms in the open $s$-scattering channel are in their ground magnetic state, so the difference $\delta\mu$ between the magnetic moments of free atoms and the bound state could only be positive. Thus, the second term in (3) is positive at the point where $a(B)=0$, which means that $a_{bg}$ is positive too.

Here, the fact that the scattering length actually crossed zero is additionally supported by the asymmetric shape of the temperature maximum. The asymmetric shape is clearly explained by the different signs of the scattering length and a sharp Feshbach resonance feature nearby. We should note that the total collisional cross-section does not hit zero at 5.83 G since there is non-vanishing anisotropic dipolar scattering cross-section [8]. Thus, we conclude that the elastic cross-section approaches its minimum at 5.83 G.

The $|a_{bg}|$ was measured by cross-dimensional rethermalization experiment [18] (see parallel PRA version for experimental details). This method allows us to measure an elastic cross-section $\sigma_{el}$, which is related to the scattering length in following way:

$$|a_{bg}| = \sqrt{\frac{\sigma_{el}}{8\pi}} \qquad (4)$$

Using (4), the background scattering length was found to be $|a_{bg}| = 144 \pm 38$ a.u..

Random to chaotic transition in the spacing of Fano-Feshbach resonances was observed in the low magnetic field Fano-Feshbach resonances in the thulium atom. The dominant role of the random statistic that presents in the formation of $s$-resonances was demonstrated. Aside from this result, the temperature shifts of different types of resonances were also studied, and radically different behaviors for $s$- and $d$- type resonances were found. The background scattering length of the thulium atom was estimated, and its positive sign was determined. The positive sign and reasonable value of its scattering length make the thulium atom a good candidate for cooling down to Bose-Einstein condensation temperatures, and its low-field Fano-Feshbach resonances make it an interesting candidate for quantum simulations.

We thank Professor Rudolf Grimm, Professor Svetlana Kotochigova and Professor Andrey Kolovsky for fruitful discussions and advises. Also we thank Aubrey Sergeant for help with manuscript preparation. This research was supported by Russian Science Foundation grant #18-12-00266.